\documentclass[RNAAS]{aastex62}

\received{February 13, 2019}
\accepted{February 18, 2019}
\submitjournal{RNAAS}

\shorttitle{BSS in NGC~2173}
\shortauthors{Dalessandro et al.}
\begin{document}

\title{The double Blue Straggler sequence in NGC~2173: yes, a field
  contamination artefact!}

\correspondingauthor{Emanuele Dalessandro}
\email{emanuele.dalessandro@inaf.it}

\author{Emanuele Dalessandro} \affil{INAF -- Astrophysics and Space
  Science Observatory Bologna, Via Gobetti 93/3 I-40129 Bologna,
  Italy}

\author{Francesco R. Ferraro} \affiliation{Universit\'a di Bologna,
  Dipartimento di Fisica e Astronomia, Via Gobetti 93/2 I-40129
  Bologna, Italy}
%\collaboration{(AAS Journals Data Scientists collaboration)}

\author{Nate Bastian} \affiliation{Astrophysics Research Institute,
  Liverpool John Moores University, 146 Brownlow Hill, Liverpool, L3
  5RF, UK}
%\affiliation{AAS Journals Associate Editor-in-Chief}
%\nocollaboration

\author{Mario Cadelano}
%\altaffiliation{Creator of AASTeX v6.2}
\affiliation{Universit\'a di Bologna, Dipartimento di Fisica e
  Astronomia, Via Gobetti 93/2 I-40129 Bologna, Italy}
%\collaboration{(LaTeX collaboration)}

\author{Barbara Lanzoni}
%\affiliation{AAS Director of Publishing}
\affiliation{Universit\'a di Bologna, Dipartimento di Fisica e
  Astronomia, Via Gobetti 93/2 I-40129 Bologna, Italy}

\author{Silvia Raso} \affiliation{Universit\'a di Bologna,
  Dipartimento di Fisica e Astronomia, Via Gobetti 93/2 I-40129
  Bologna, Italy}
%\affiliation{IOP Publishing, Washington, DC 20005}



%\begin{abstract}

%\end{abstract}



\section{Introduction} \label{sec:intro}
\citet[hereafter L18a]{li18a} claimed that the
young stellar cluster NGC~2173 in the Large Magellanic Cloud (LMC)
harbours a bifurcated sequence of blue straggler stars (BSSs), similar
to those detected in a few dynamically old globular clusters
\citep{ferraro09,dalessandro13,simunovic14}. However,
\citet[hereafter D19]{dalessandro19} re-analyzed the data by
taking into account the contamination of the cluster population from
LMC field stars, which was completely neglected by L18a.  D19
showed that $\sim 40\%$ of the selected BSS sample (and especially the
population observed along one of the two sequences) is composed of
field star interlopers, concluding that the double BSS sequence is
most likely a {\it field contamination artefact}.

In a recent note \citet[hereafter L18b]{li18b} argued that the analysis by 
D19 is affected by two issues related to (1) the use of different HST
instruments/filters in the decontamination procedure,
and (2) a presumed overestimate of the number of field stars, 
which show a central radial segregation.

Here we demonstrate that the D19 decontamination procedure is
completely unaffected by the use of different HST
instruments, and that, the field stars removed by D19 have no significant radial
segregation toward the cluster center. Hence, we confirm that the
claimed double sequence is just a field contamination artefact.

\begin{figure}
\centering \includegraphics[width=12cm]{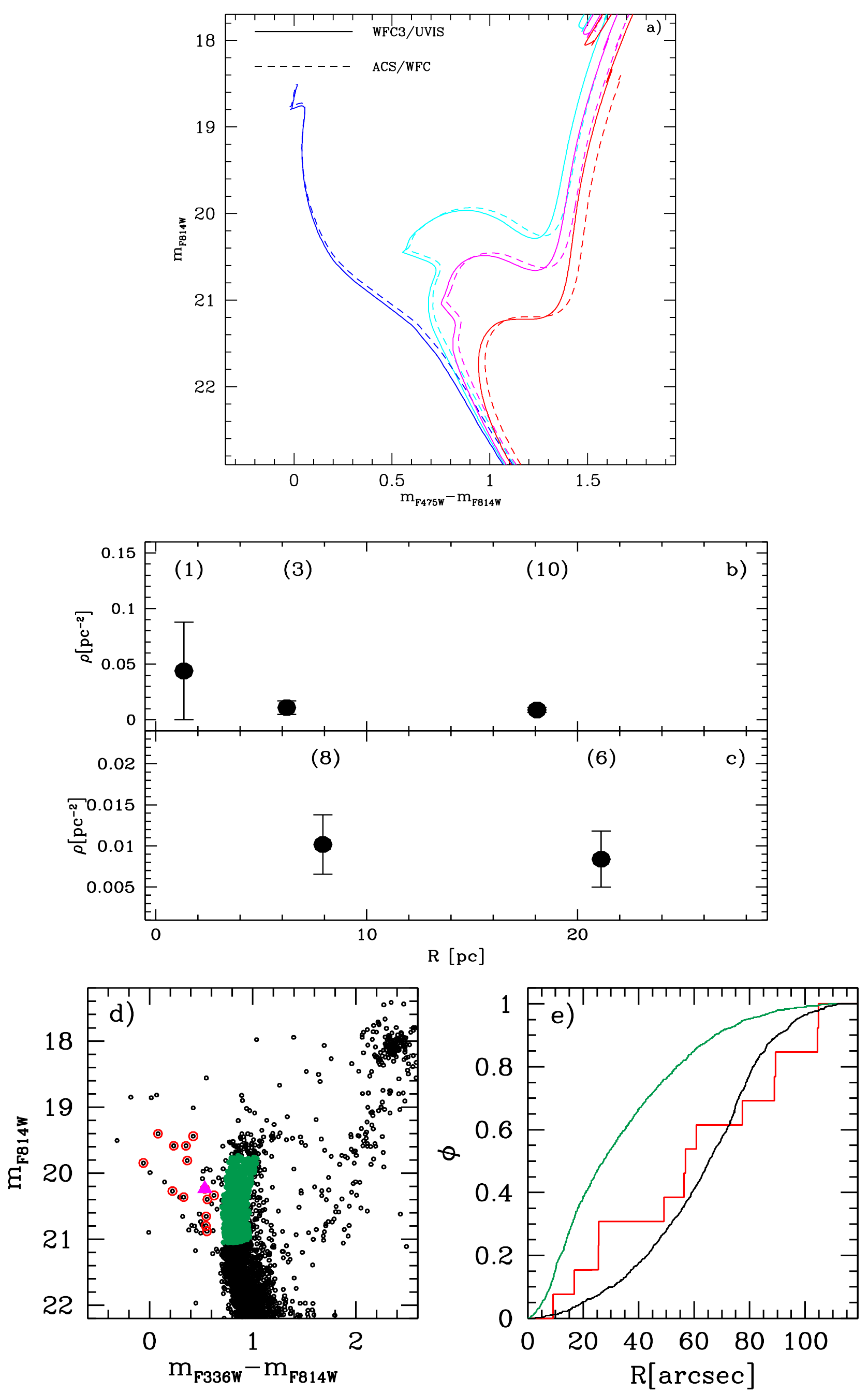}
\caption{{\bf a)}: Comparison between WFC3/UVIS (solid lines) and ACS/WFC (dashed) 
isochrones \citep{pietrinferni06} with ages of 0.4
  Gyr (blue), 2.0 Gyr (cyan), 6.0 Gyr (magenta) and 8.0 Gyr (red) and
  metallicity [Fe/H]$\sim-0.5$ in the ($m_{F814W}$, $m_{F475W}$-$m_{F814W}$) CMD.
  {\bf b)}: Density distribution of the D19 sample of (14) removed BSSs obtained with
  the same radial bins adopted by Li18b and {\bf c)}, by dividing
  the field of view in two concentric radial regions covering the same
  area.  The number of stars observed in the each
  radial bin is marked within brackets.  {\bf d)}: CMD of NGC~2173
  zoomed in the MS turn-off region. The red circles are the stars removed by D19 as likely field contaminants
  of the BSS population, the magenta triangle
  corresponds to the only object removed at $r<r_c$. Green circles
  mark the MS population selected as reference.  {\bf e)}: The red
  and green lines show the cumulative radial distribution of the
  populations selected in panel d) with the same color code. The black
  line refers to a mock population homogeneously distributed within
  the WFC3 area and representative of the LMC field population.}
\end{figure}

\begin{figure}
\centering \includegraphics[width=14cm]{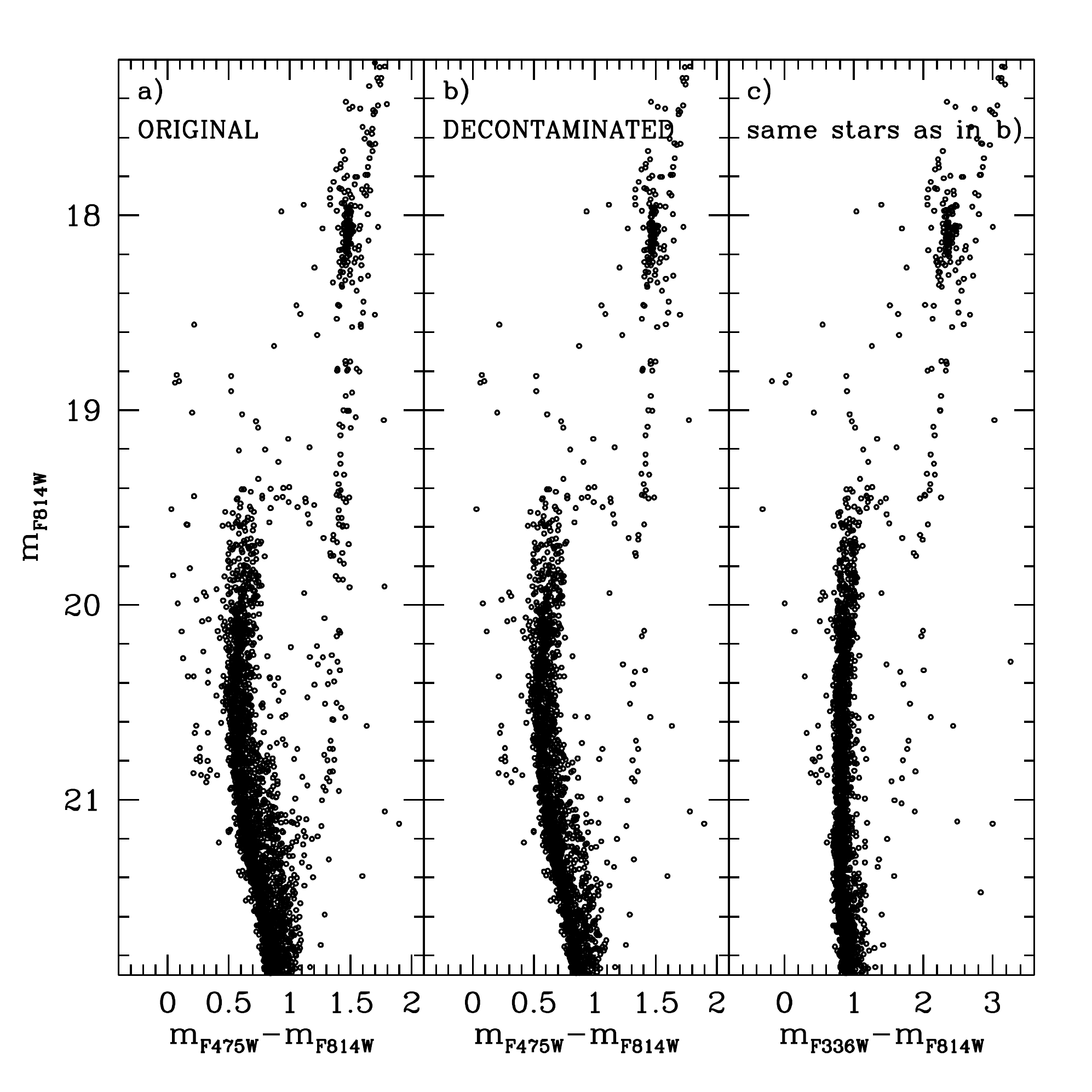}
\caption{{\bf a)}: original ($m_{\rm F814W}, m_{\rm F475W}-m_{\rm F814W}$) CMD;
{\bf b)}: CMD of stars surviving to the statistical decontamination analysis described in D19;
{\bf c)}: ($m_{\rm F814W}, m_{\rm F336W}-m_{\rm F814W}$) CMD of the same stars as in panel b).}
\end{figure}

\section{Discussion}
For the field decontamination procedure, D19 made use of the {\it
  Cluster} and the {\it Field Pointings}, secured in the same set of
filters (F475W and F814W) with the WFC3/UVIS and the ACS/WFC,
respectively. Figure~1a shows that the filter systems of these two
instruments are essentially the same.  Indeed, \citet{deustua18} report differences 
smaller than $\pm 0.02$ mag  (in a wide color range)
between the ACS/WFC and WFC3/UVIS F814W and F475W magnitudes.
Hence, for their
field decontamination analysis, D19 did not ``cross-calibrate the
instrumental magnitudes difference between two different filter
systems'', at odds with what is stated by L18b.  On the same line, we stress that the different
main sequence (MS) slopes observed in the two HST data sets (see
Figure~4 in D19) is simply due to different ages and age distributions of the
cluster and field populations. 
The stars that survived the
decontamination procedure (performed in the $m_{F475W}-m_{F814W}$ diagram), were 
then analysed in the ($m_{\rm F814W},m_{\rm
  F336W}-m_{\rm F814W}$) color-magnitude diagram (CMD -- see Figure~2), where the
presence of a double BSS sequence was claimed by L18a. This
is straightforward since every star in the {\it Cluster Pointing} has also
a measure in the F336W (see Table 1 in D19).

\begin{figure}
\centering \includegraphics[width=16cm]{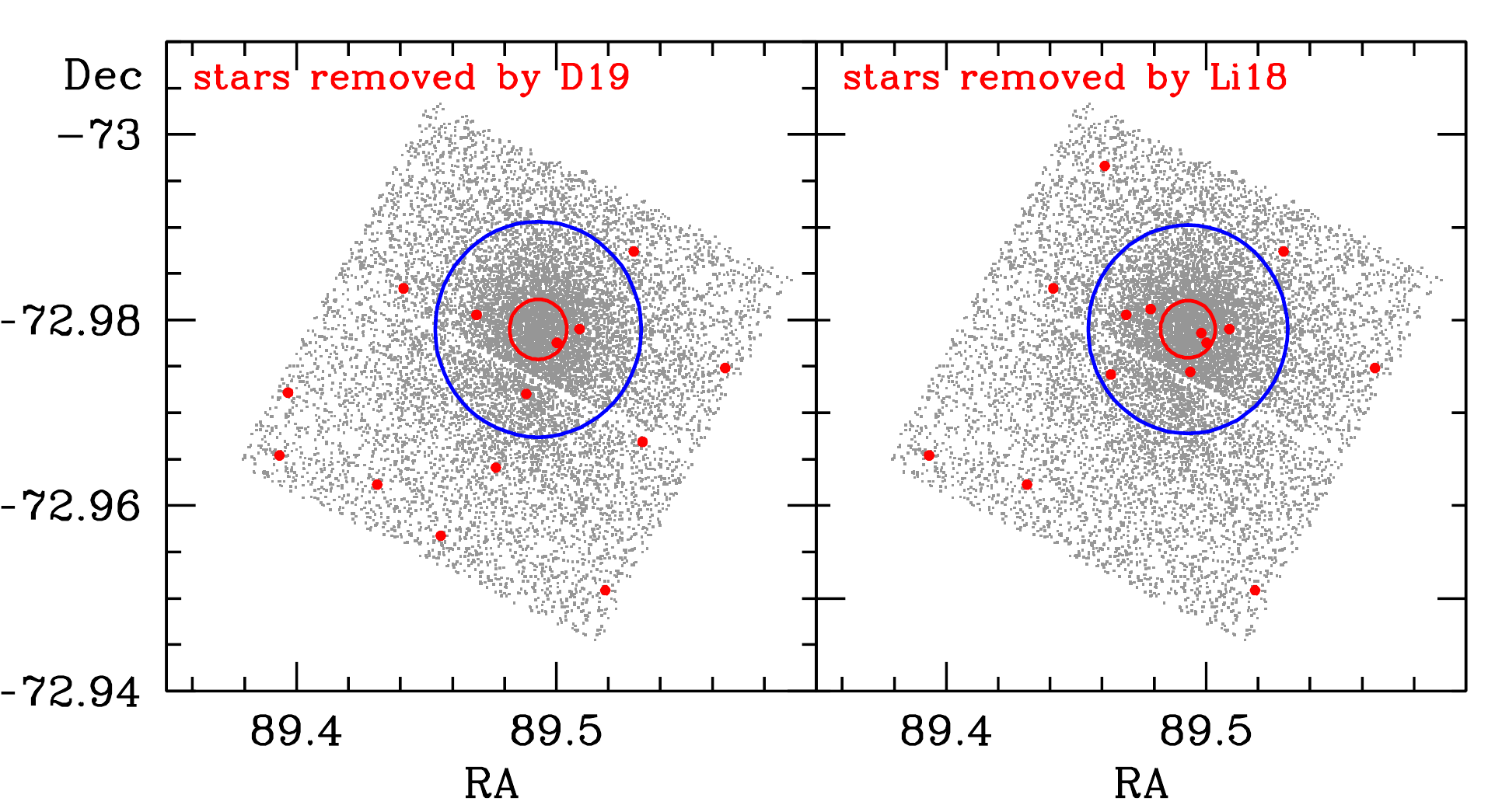}
\caption{Comparison between the distribution of the BSS field stars removed by D19 (left panel) and 
those erroneously identified/selected by L18b as the``D19 removed stars" (right panel). 
Clearly most of them are not the same.}
\end{figure}

As for the presumed radial segregation of the D19 field contaminants,
we first have to warn that the 14 objects selected by L18b as the
``D19 field stars'' are not the same used in D19 (see Figure~3). 
In Figure~1b we plot the radial density
distribution obtained by using the true D19 sample of removed stars,
and the same radial bins adopted in L18b.  
While a quick look to the plot
reveals an apparent overdensity at the center, we stress that it
is caused by only one (!!)  star.
Very basic statistical arguments demonstrate that (because of the
very poor statistics) the density is constant
when Poissonian errors are taken into account: $\rho_1=0.044\pm0.044$,
$\rho_2=0.011\pm0.006$, $\rho_3=0.009\pm0.002$, in units of stars
pc$^{-2}$.  
%We also stress that in more than
%$80\%$ of all the decontamination runs performed 
%in D19, no BSS are subtracted for $r<r_c$.  
To quantify the probability that the density distribution shown in in Figure~1b is obtained by
chance,
we run 10000 Monte Carlo simulations extracting, each time,
14 objects homogeneously distributed within the WFC3 field of view and
counting the number of stars in the three adopted radial bins.
We find that in more than $99.9\%$ of the cases $1\pm1$ star is counted at $r<r_c$, 
thus demonstrating that the apparent central peak hasn't any
statistical significance. 
In addition, we stress that no stars are subtracted at $r<r_c$ in $80\%$ 
of all the decontamination runs performed in D19. 
We also note that if the sample is split in
two concentric radial regions covering the same area, the derived BSS
densities (Figure~1c) are indistinguishable ($\rho_1=0.010\pm0.004$ pc$^{-2}$
and $\rho_2=0.008\pm0.003$ pc$^{-2}$), as expected in the case of a uniform
spatial distribution, as that of the field population.  The definitive
demonstration that the sample of stars removed by D19 is effectively
composed of likely field contaminants is shown in Figure~1e.  As
apparent, the cumulative radial distribution of the 14 removed stars
(red circles in Figure~1d) is inconsistent (KS-test
probability$\sim0.1\%$) with the cluster radial distribution (as
traced by the MS stars, in green), while it is well compatible
(KS-test probability$\sim60\%$) with that of a mock population of
1000 stars homogeneously distributed within the WFC3 field of view (in
black), which is representative of the LMC field.

\begin{figure}
\centering \includegraphics[width=14cm]{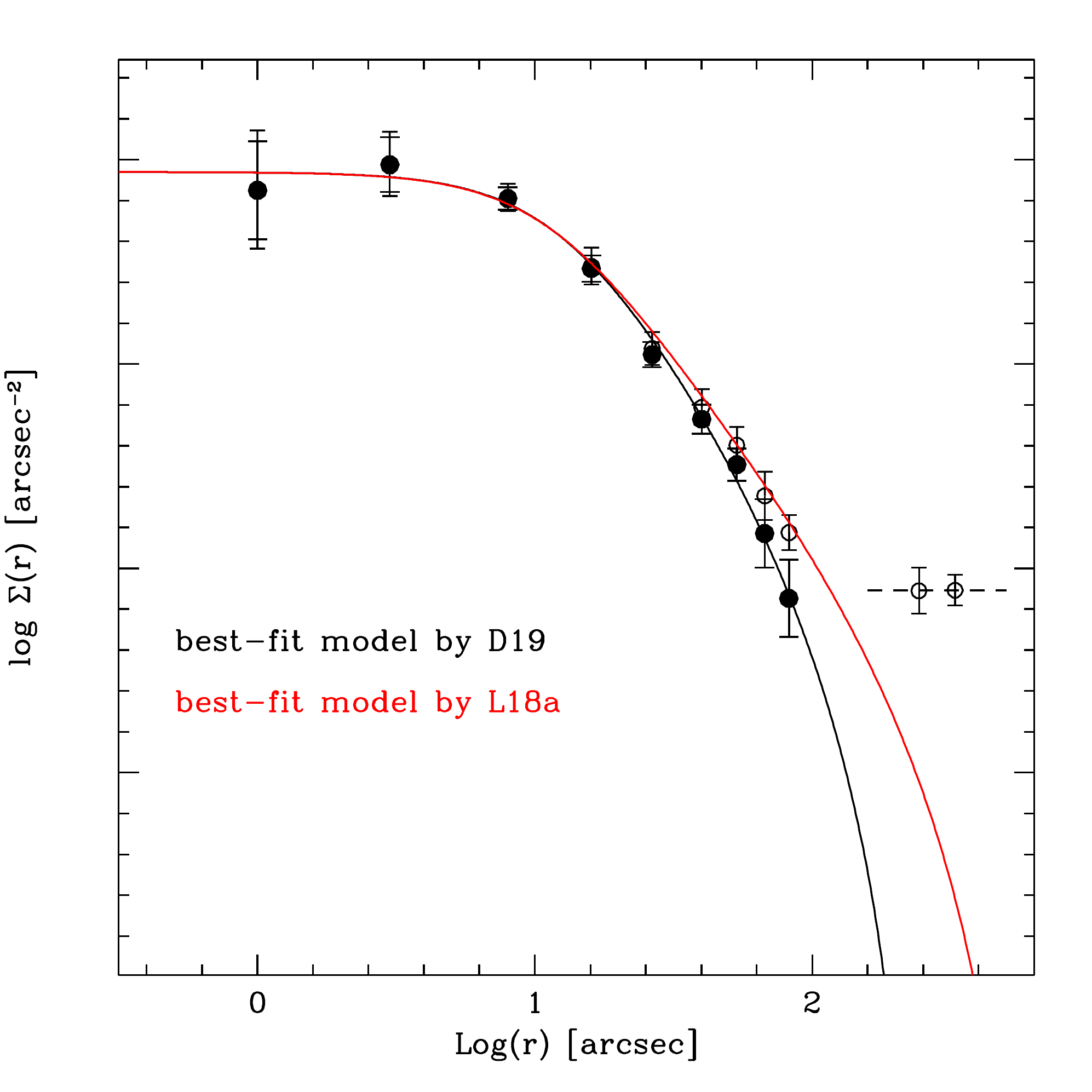}
\caption{Observed and decontaminated star count density profile of NGC~2173 (the same as 
in Figure~3 of D19).
The dashed line represents the density value of the background as derived by averaging the 
two outermost radial bins, which have been obtained from the {\it Field Pointing}.
The best-fit King model obtained by D19 is shown in black, while the the model 
corresponding to the structural parameters quoted by L18a is shown in red.
Clearly the red model fits the non-background subtracted profile while it is unable to reproduce 
the cluster density distribution properly corrected for the LMC field contamination.}
\end{figure}

Clearly, these results do not change even if a $10\%$ incompleteness is
assumed within $r_c$ (as estimated by L18a).   

As a final note, we stress that the density
profile obtained from the {\it Field Pointing} remains constant with
radius (see empty circles at $r> 200\arcsec$ in Figure~4)
as expected for field stars, thus
fully justifying the use of the {\it Field Pointing}
for decontamination purposes. Figure~4 shows that the structural parameters estimated by
L18a well fit the observed density distribution if no
subtraction of the field contribution is performed, 
thus causing a significant overestimate of the
cluster tidal radius.

\section{Conclusion}
These results demonstrate that most of the
14 stars removed by D19 are indeed field interlopers.  Note that even
by conservatively assuming that the star removed in the central bin
(magenta triangle in Figure~1d) is a cluster member, the BSS region in the CMD of
NGC~2173 remains populated by only one sequence.  Therefore the
conclusions presented in D19 remain unchanged:

{\it (i)} The {\it Field Pointing} is perfectly suitable 
to estimate the field star contamination of NGC 2173.

{\it (ii)} The D19 analysis does not overestimate the BSS contamination level.

{\it (iii)} The presumed double BSS sequence in NGC~2173 is a
contamination artefact, since when field contamination is properly
taken into account, only a single and scarcely populated BSS sequence
appears.



\end{document}